\documentclass{bmcart}

\usepackage[utf8]{inputenc} 
\usepackage{xspace}
\usepackage{amsmath}
\usepackage[]{subcaption}
\usepackage{graphicx}
\usepackage{multirow}
\usepackage{tikz}

\startlocaldefs

\newcommand{\etc}{et al.}
\newcommand{\covid}{COVID-19\xspace}
\newcommand{\method}{Flight-SEIR\xspace}
\newcommand{\popr}{$R0$\xspace}
\newcommand{\popri}{$R0_i$\xspace}

\endlocaldefs

\begin{document}

\newcounter{observation}[section]
\newenvironment{observation}[1][]{\refstepcounter{observation}\par\medskip
   \textbf{Observation~\theobservation:  #1} \itshape}{\medskip}

\begin{frontmatter}

\begin{fmbox}
\dochead{Research}


\title{Incorporating Dynamic Flight Network in SEIR to Model Mobility between Populations } 

\author[
   addressref={aff1,aff2},                   
   corref={aff1,aff2},                       
   email={xiaoye.ding@mail.mcgill.ca}   
]{Xiaoye Ding}

\author[
   addressref={aff1,aff2},
   email={shenyang.huang@mail.mcgill.ca}
]{Shenyang Huang}

\author[
   addressref={aff1,aff2},
   email={oi.k.leung@mail.mcgill.ca}
]{Abby Leung}

\author[
   addressref={aff1,aff2},                   
   noteref={chair},                        
   email={rrabba@cs.mcgill.ca}   
]{Reihaneh Rabbany}


\address[id=aff1]{
  \orgname{School of Computer Science, McGill University}, 
  \city{Montreal},                              
  \cny{Canada}                                    
}

\address[id=aff2]{%
  \orgname{Mila, Quebec Artificial Intelligence Institute},
  \city{Montreal},
  \cny{Canada}
}


\begin{artnotes}
\note[id=chair]{CIFAR AI chair} 
\end{artnotes}

\end{fmbox}


\begin{abstractbox}

\begin{abstract} 

Current efforts of modelling COVID-19 are often based on the standard compartmental models such as SEIR and their variations. As pre-symptomatic and asymptomatic cases can spread the disease between populations through travel, it is important to incorporate mobility between populations into the epidemiological modelling. 
In this work, we propose to modify the commonly-used SEIR model to account for the dynamic flight network, by estimating the imported cases based on the air traffic volume as well as the test positive rate at the source. This modification, called \method, can potentially enable 1). early detection of outbreaks due to imported pre-symptomatic and asymptomatic cases, 2). more accurate estimation of the reproduction number and 3). evaluation of the impact of travel restrictions and the implications of lifting these measures. The proposed \method is essential in navigating through this pandemic and the next ones, given how interconnected our world has become. 

\end{abstract}




\end{abstractbox}
%

\end{frontmatter}

\section{Introduction}

The coronavirus disease 2019 (COVID-19) pandemic spread rapidly around the world and has affected countless lives. As of July 4, there have been 10,922,324 confirmed cases and 523,011 deaths globally~\cite{whosr166}. Of all places, but the origin, the start of the spread is related to travel. For example, in Canada, the first travel-related case was identified in January 2020 and by April 2020, community transmission was present in almost all provinces~\cite{ogden2020modelling}. To better understand and effectively model a global pandemic such as COVID-19, it is important to consider the following questions: 1). \emph{how to model the spread of a disease from one community to another}, 2). \emph{how to estimate the likelihood of an outbreak caused by imported cases} and 3) \emph{how to measure the effectiveness of travel bans and evaluate the impact of lifting those restrictions}. In this work, we aim to answer the above questions by incorporating dynamic flight network into epidemiological modelling.

\begin{figure}[h!]
     \centering
        \begin{subfigure}[b]{.95\textwidth}
        \includegraphics[width=1\textwidth]{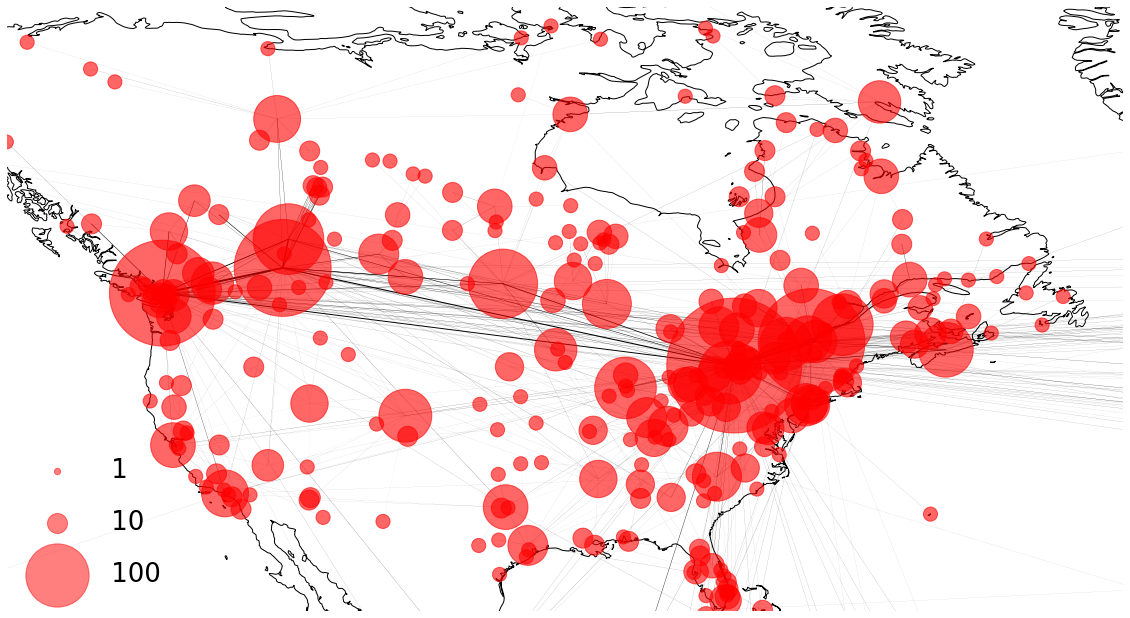}
        \caption{\csentence{Flight Network on January 20th, 2020} }   \label{fig:fn1}
      \end{subfigure}
    \begin{subfigure}[b]{.95\textwidth}
        \includegraphics[width=1\textwidth]{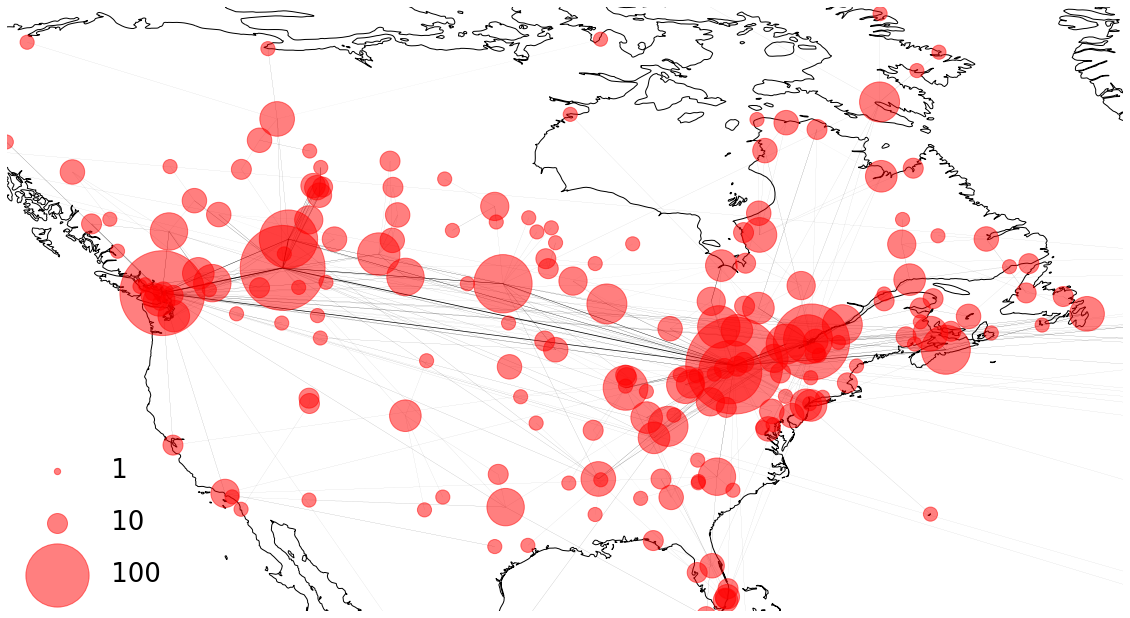}
        \caption{\csentence{Flight Network on April 2nd, 2020}}\label{fig:fn2}
      \end{subfigure}
 \caption{\csentence{Flight network before and after imposing travel restrictions.}\\Note that only flights with an endpoint in Canada have been considered.}
     \label{fig:fligthOverview}
\end{figure}

Most models view each population as a closed system, isolated from the rest of the world. However, this assumption is rarely true in practice. For example, in Canada, even with the travel restrictions in place, there are still a considerable amount of flights going in and out of the country. As shown in Figure \ref{fig:fligthOverview}, while there was a significant amount of reduction in daily flights to and from Canada, much air traffic remained well after the travel restrictions were put in place, i.e. March 16th with countries excluding US and March 20th with US. Here, each node represents one of the airports in North America. Both the size of the node and the thickness of the edge are in proportion to the volume of flights. Note that we focus on flights data arriving at or departed from Canada, which is the population under study for this work. The node size of US airports only reflect their traffic volumes related to Canada. 


To account for mobility between populations, we propose \method, a modification of the standard SEIR model to include the mobility of exposed individuals between the target population and the rest of the world. We show how \method is able to raise alarms of upcoming outbreaks even when there is no community transmission. We believe that \method is crucial for alerting public agencies ahead of an outbreak similar to \covid. Moreover, we show that \method is able to fit the real data more closely and provide a better estimation of how contagious the disease is when compared to the standard SEIR model which either underestimates or overestimates the size of infectious population by assuming a closed population. In the context of \covid, there is an immense political and economical pressure for the governments to reopen the borders and resume flights. \method can inform policy makers by predicting the epidemic curves with projected flight data therefore performing risk estimation of lifting air travel restrictions. 


\section{Related Works}

Many recent works model COVID-19 with variants of the classical susceptible, infected and recovered~(SIR) and the susceptible, exposed, infected and recovered~(SEIR) model~\cite{katul2020global, ogden2020modelling, tuite2020mathematical, linka2020reproduction}. The focus of these studies is often placed on the estimation of the basic reproduction number $R_0$~\cite{katul2020global}, the average number of new infections an infected individual causes in a fully susceptible population~\cite{barabasi2013network}. 

One key challenge of modelling COVID-19 is the existence of asymptomatic and pre-symptomatic transmission~\cite{wei2020presymptomatic,vetter2020clinical}. These pre-symptomatic and asymptomatic carriers of COVID-19 can travel to a foreign country and initiate the spread of COVID-19 which eventually leads to an outbreak. 
Motivated by this, some studies consider the travel between populations. For example, Bogoch \etc~\cite{bogoch2020potential} estimated the risk of an outbreak by presenting the Infectious Disease Vulnerability Index (IDVI) for international destinations receiving the highest number of passengers from major cities in China. Kraemer \etc\cite{kraemer2020effect} and Linka \etc~\cite{linka2020reproduction} studied the correlation between human mobility and the spread of the disease. These models are, however, not epidemiological models and cannot project the spread of disease based on travel patterns. 
 
Lin \etc~\cite{lin2020conceptual} introduced a step-wise emigration rate to model the before and after of Wuhan lock down. Yang \etc~\cite{yang2020modified} considered a dynamic population with inflow and outflow of susceptible and exposed individuals between different provinces in China. The size of inflow and outflow depends on the migration index, defined based on the number of inbound and outbound events by rail, air and road traffic. Yamana \etc~\cite{yamana2020projection} assessed the impact of reopening states in the US by considering the daily work movements and random movements between counties. However, these models only consider disease spread within the same country. In comparison, our method also incorporates international flights to effectively understand how \covid spreads globally. 

Some models do account for the global spread of \covid. Kucharski \etc~\cite{kucharski2020early} studied domestic cases within Wuhan and international cases that originated from Wuhan. In particular, they modelled how individuals that are exposed to \covid in Wuhan can spread the disease to other countries. They assumed that the probability of cases being exported from Wuhan depended on the number of cases in Wuhan, the number of outbound travellers, the relative connectivity of different countries and the relative probability of reporting a case in each country. Wu \etc~\cite{wu2020nowcasting} inferred \popr and the outbreak size by estimating the number of cases exported from Wuhan. They also forecasted the spread of the disease within and outside Wuhan by accounting for mobility and transmissibility reduction. Chinazzi \etc\cite{chinazzi2020effect} applied the Global Epidemic and Mobility Model to simulate the effect of travel ban in China. International mobility is set to reduce by 40\% and 90\% after the travel ban while domestic mobility is derived from mobility indices. These models either assume that the global travel behavior remains the same throughout the pandemic or used a step-wise function to model mobility patterns before and after international travel restrictions. In contrast, \method incorporates dynamic flight network which models more granular changes in the air traffic and provides the flexibility of plugging in projected flight network to simulate different levels of travel restrictions. In addition, we are not aware of any work that studies how resuming flights will affect the spread of \covid, which \method provides.

While travel-related Non-Pharmaceutical Interventions~(NPIs) played a significant role in confining \covid, some studies focused on other NPIs such as social distancing and quarantine~\cite{tuite2020mathematical, ogden2020predictive, ferguson2020report, block2020social, reich2020modeling}. For instance, Ogden \etc~\cite{ogden2020predictive} described the predictive modelling efforts for COVID-19 within the Public Health Agency of Canada. They modelled the effects of physical distancing by reducing daily per capita contact rates. A separate agent model is used to simulate the effects of closing schools, workplaces and other public places. \cite{ferguson2020report} employed an individual-based simulation model to evaluate the impact of NPIs, such as quarantine, social distancing and school closure. \cite{reich2020modeling} compared the impact of testing, contact tracing, quarantine and social distancing with a network-based epidemic model.  

Another key challenge in modelling COVID-19 is to account for various biases such as limitations in testing capacity and individually varied lag time between infection and final clinical outcome. Verity \etc~\cite{verity2020estimates} provided an estimate for the severity of COVID-19 which corrected for censoring and ascertainment biases. They argued that the crude case fatality ratio obtained by dividing the number of death by the number of cases is misleading. In particular, during the beginning phase of an epidemic, the diagnostic capacity is low and surveillance is typically biased towards detecting clinically severe cases. Abdollahi \etc~\cite{abdollahi2020temporal} made a similar argument while focusing on COVID-19 spread in Canada and US. Limited ability to test or recognize mildly or moderately symptomatic people in US and Canada has likely led to substantial underestimation of the infection rate. \method provides a way to mitigate the various biases presented in the early stage of a pandemic by leveraging the international flight network and estimating how many exposed individuals are imported through flights. This risk estimation technique can be applied even when the testing capacity of a region is limited or biased towards clinically severe cases. 

\begin{figure}
    \centering
    \resizebox{0.4\textwidth}{!}{
    \tikzset{every picture/.style={line width=0.75pt}} 

\begin{tikzpicture}[x=0.75pt,y=0.75pt,yscale=-1,xscale=1]

\draw  [fill={rgb, 255:red, 155; green, 155; blue, 155 }  ,fill opacity=0.15 ] (156,142.36) .. controls (156,84.96) and (168.9,38.43) .. (184.82,38.43) .. controls (200.74,38.43) and (213.64,84.96) .. (213.64,142.36) .. controls (213.64,199.76) and (200.74,246.29) .. (184.82,246.29) .. controls (168.9,246.29) and (156,199.76) .. (156,142.36) -- cycle ;
\draw  [fill={rgb, 255:red, 184; green, 233; blue, 134 }  ,fill opacity=0.61 ] (169.36,67.71) -- (197.3,67.71) -- (197.3,92.04) -- (169.36,92.04) -- cycle ;
\draw  [fill={rgb, 255:red, 251; green, 233; blue, 18 }  ,fill opacity=0.37 ] (169.45,109.52) -- (197.4,109.52) -- (197.4,133.84) -- (169.45,133.84) -- cycle ;

\draw  [fill={rgb, 255:red, 244; green, 139; blue, 139 }  ,fill opacity=0.67 ] (170.08,151.63) -- (198.02,151.63) -- (198.02,175.95) -- (170.08,175.95) -- cycle ;

\draw  [fill={rgb, 255:red, 156; green, 136; blue, 118 }  ,fill opacity=0.41 ] (170.18,193.44) -- (198.12,193.44) -- (198.12,217.76) -- (170.18,217.76) -- cycle ;

\draw    (182.48,93.85) -- (182.48,106.39) ;
\draw [shift={(182.48,108.39)}, rotate = 270] [color={rgb, 255:red, 0; green, 0; blue, 0 }  ][line width=0.75]    (10.93,-3.29) .. controls (6.95,-1.4) and (3.31,-0.3) .. (0,0) .. controls (3.31,0.3) and (6.95,1.4) .. (10.93,3.29)   ;
\draw  [fill={rgb, 255:red, 155; green, 155; blue, 155 }  ,fill opacity=0.38 ] (93.93,54.07) .. controls (93.93,45.35) and (100.34,38.29) .. (108.25,38.29) .. controls (116.16,38.29) and (122.57,45.35) .. (122.57,54.07) .. controls (122.57,62.78) and (116.16,69.85) .. (108.25,69.85) .. controls (100.34,69.85) and (93.93,62.78) .. (93.93,54.07) -- cycle ;
\draw  [fill={rgb, 255:red, 155; green, 155; blue, 155 }  ,fill opacity=0.38 ] (79.93,125.21) .. controls (79.93,116.5) and (86.34,109.43) .. (94.25,109.43) .. controls (102.16,109.43) and (108.57,116.5) .. (108.57,125.21) .. controls (108.57,133.93) and (102.16,141) .. (94.25,141) .. controls (86.34,141) and (79.93,133.93) .. (79.93,125.21) -- cycle ;
\draw    (182.48,134.75) -- (182.48,147.29) ;
\draw [shift={(182.48,149.29)}, rotate = 270] [color={rgb, 255:red, 0; green, 0; blue, 0 }  ][line width=0.75]    (10.93,-3.29) .. controls (6.95,-1.4) and (3.31,-0.3) .. (0,0) .. controls (3.31,0.3) and (6.95,1.4) .. (10.93,3.29)   ;
\draw    (183.41,176.56) -- (183.41,189.1) ;
\draw [shift={(183.41,191.1)}, rotate = 270] [color={rgb, 255:red, 0; green, 0; blue, 0 }  ][line width=0.75]    (10.93,-3.29) .. controls (6.95,-1.4) and (3.31,-0.3) .. (0,0) .. controls (3.31,0.3) and (6.95,1.4) .. (10.93,3.29)   ;
\draw  [fill={rgb, 255:red, 155; green, 155; blue, 155 }  ,fill opacity=0.38 ] (91.93,225.21) .. controls (91.93,216.5) and (98.34,209.43) .. (106.25,209.43) .. controls (114.16,209.43) and (120.57,216.5) .. (120.57,225.21) .. controls (120.57,233.93) and (114.16,241) .. (106.25,241) .. controls (98.34,241) and (91.93,233.93) .. (91.93,225.21) -- cycle ;
\draw  [fill={rgb, 255:red, 155; green, 155; blue, 155 }  ,fill opacity=0.38 ] (87.93,168.7) .. controls (87.93,167.36) and (88.91,166.29) .. (90.11,166.29) .. controls (91.32,166.29) and (92.3,167.36) .. (92.3,168.7) .. controls (92.3,170.03) and (91.32,171.11) .. (90.11,171.11) .. controls (88.91,171.11) and (87.93,170.03) .. (87.93,168.7) -- cycle ;
\draw  [fill={rgb, 255:red, 155; green, 155; blue, 155 }  ,fill opacity=0.38 ] (88.38,176.64) .. controls (88.38,175.31) and (89.36,174.23) .. (90.56,174.23) .. controls (91.77,174.23) and (92.74,175.31) .. (92.74,176.64) .. controls (92.74,177.97) and (91.77,179.05) .. (90.56,179.05) .. controls (89.36,179.05) and (88.38,177.97) .. (88.38,176.64) -- cycle ;
\draw  [fill={rgb, 255:red, 155; green, 155; blue, 155 }  ,fill opacity=0.38 ] (89.28,184.58) .. controls (89.28,183.25) and (90.26,182.17) .. (91.46,182.17) .. controls (92.67,182.17) and (93.64,183.25) .. (93.64,184.58) .. controls (93.64,185.92) and (92.67,187) .. (91.46,187) .. controls (90.26,187) and (89.28,185.92) .. (89.28,184.58) -- cycle ;
\draw    (120.64,47.29) .. controls (140.64,56.29) and (157.64,68.29) .. (169.45,109.52) ;
\draw [shift={(152.93,72.39)}, rotate = 232.25] [fill={rgb, 255:red, 0; green, 0; blue, 0 }  ][line width=0.08]  [draw opacity=0] (10.72,-5.15) -- (0,0) -- (10.72,5.15) -- (7.12,0) -- cycle    ;
\draw    (107.64,120.29) .. controls (125.64,117.29) and (144.64,113.29) .. (169.64,117.29) ;
\draw [shift={(138.77,115.95)}, rotate = 534.98] [fill={rgb, 255:red, 0; green, 0; blue, 0 }  ][line width=0.08]  [draw opacity=0] (10.72,-5.15) -- (0,0) -- (10.72,5.15) -- (7.12,0) -- cycle    ;
\draw    (112.64,211.29) .. controls (121.64,187.29) and (144.64,157.29) .. (169.45,133.84) ;
\draw [shift={(137.24,169.54)}, rotate = 486.72] [fill={rgb, 255:red, 0; green, 0; blue, 0 }  ][line width=0.08]  [draw opacity=0] (10.72,-5.15) -- (0,0) -- (10.72,5.15) -- (7.12,0) -- cycle    ;
\draw    (169.45,109.52) .. controls (141.64,85.29) and (136.64,73.29) .. (121.64,61.29) ;
\draw [shift={(145.15,85.78)}, rotate = 407.28] [fill={rgb, 255:red, 0; green, 0; blue, 0 }  ][line width=0.08]  [draw opacity=0] (10.72,-5.15) -- (0,0) -- (10.72,5.15) -- (7.12,0) -- cycle    ;
\draw    (169.64,120.29) .. controls (142.64,131.29) and (134.64,131.29) .. (108.64,127.29) ;
\draw [shift={(139.54,129.33)}, rotate = 350.68] [fill={rgb, 255:red, 0; green, 0; blue, 0 }  ][line width=0.08]  [draw opacity=0] (10.72,-5.15) -- (0,0) -- (10.72,5.15) -- (7.12,0) -- cycle    ;
\draw    (169.45,133.84) .. controls (159.64,165.29) and (137.64,193.29) .. (120.64,219.29) ;
\draw [shift={(147.74,178.61)}, rotate = 301.48] [fill={rgb, 255:red, 0; green, 0; blue, 0 }  ][line width=0.08]  [draw opacity=0] (10.72,-5.15) -- (0,0) -- (10.72,5.15) -- (7.12,0) -- cycle    ;

\draw (177.69,69.74+5) node [anchor=north west][inner sep=0.75pt]    {$S_i$};
\draw (186.04,87.92+7) node [anchor=north west][inner sep=0.75pt]    {$\beta_i $};
\draw (186.04,130.63+7) node [anchor=north west][inner sep=0.75pt]    {$\sigma $};
\draw (186.04,169.71+7) node [anchor=north west][inner sep=0.75pt]    {$\gamma $};
\draw (97,42+7) node [anchor=north west][inner sep=0.75pt]    {$C_{1}$};
\draw (83,113+7) node [anchor=north west][inner sep=0.75pt]    {$C_{2}$};
\draw (95,213+7) node [anchor=north west][inner sep=0.75pt]    {$C_{n}$};
\draw (131,37+5) node [anchor=north west][inner sep=0.75pt]  [font=\small]  {$E^{in}_{1}$};
\draw (107,94+7) node [anchor=north west][inner sep=0.75pt]  [font=\small]  {$E^{in}_{2}$};
\draw (102,169+7) node [anchor=north west][inner sep=0.75pt]  [font=\small]  {$E^{in}_{n}$};
\draw (130,193+7) node [anchor=north west][inner sep=0.75pt]  [font=\small]  {$E^{out}_{n}$};
\draw (108.64,127.29+5) node [anchor=north west][inner sep=0.75pt]  [font=\small]  {$E^{out}_{2}$};
\draw (108.25,69.85+5) node [anchor=north west][inner sep=0.75pt]  [font=\small]  {$E^{out}_{1}$};
\draw (175.84,194.25+7) node [anchor=north west][inner sep=0.75pt]    {$R_i$};
\draw (178,151.54+7) node [anchor=north west][inner sep=0.75pt]    {$I_i$};
\draw (175.84,109.73+7) node [anchor=north west][inner sep=0.75pt]    {$E_i$};

\end{tikzpicture}

    }
    \caption{\csentence{Demographic and epidemic dynamics of \method.\\} The movements of exposed individuals between the populations are modelled. \\ An exposed individual can either come from other countries or be infected by \\ an infectious individual within the same population.}
    \label{fig:abstract}
\end{figure}

\section{\method}

\begin{table}[t]
    \small
    \begin{tabular}{c|l|c|c}
        \hline
        Parameter & Description & Value & Type \\
        \hline
        $S_i$ & \# susceptible individuals at node $i$ & & \\
        $E_i$ &  \# exposed individuals at node $i$ & & \\
        $E_{ij}$ &  \# exposed individuals travelling from node $i$ to node $j$ & & \\
        $E_{ji}$ &  \# exposed individuals travelling from node $j$ to node $i$ & & \\
        $I_i$ &  \# infected individuals at node $i$ & & \\
        $R_i$ &  \# recovered individuals at node $i$ & & \\
        $N_i$ &  total population at node $i$ & & constant \\
        $C$ & mean latent period of the disease & 5 & constant\\
        $\sigma$ & incubation rate $E_i \rightarrow I_i$ & $\frac{1}{C}$ & constant\\
        $D$ & mean infectious period & 14 & constant\\
        $\gamma$ & recovery rate $I_i \rightarrow R_i$ & $\frac{1}{D}$ & constant\\
        $\beta_{i}$ & transmission rate $S_i \rightarrow E_i$ for population $i$ & & fitted\\
        \popri & the population reproduction number  & $\frac{\beta_{i}}{\gamma}$& fitted\\
        $Passengers_{ij}$ & \# passengers travelling from node $i$ to $j$ & & estimated\\
        $Passengers_{ji}$ & \# passengers travelling from node $j$ to $i$ & & estimated\\
        $P_i$ & test positive rate of node $i$ & & estimated\\
        $\eta$ & \% of exposed individuals over all infected individuals & & estimated\\
        $F_{ij}$ & \# flights from node $i$ to node $j$ & & estimated \\
        $CAP$ & estimated average flight capacity & & estimated\\
        $LF$ & load factor: onboard passengers to available seats ratio & & estimated\\
        $\tau$ & \% of projected air traffic over pre-pandemic air traffic& & variable\\
        \hline
    \end{tabular}
    \caption{ \textbf{Parameters for \method.} Type of the parameter can be constant, variable, fitted or estimated. Constant parameters are set based on~\cite{whocjm, CanPop}.}
    \label{tab:param}
\end{table}

We modified the differential equations of a standard SEIR model to include demographics dynamics derived from the flight network. This modified SEIR model is called \method. More specifically, we consider an open population setting where exposed individuals~(the $E$ compartment) are free to travel in and out of the population as it is difficult to determine if they have been infected or not. For simplicity, we assume that individuals who are in the infected compartment, $I$, would be denied boarding flights and therefore cannot travel between populations. Figure~\ref{fig:abstract} illustrates the epidemic and demographic dynamics modelled by \method, and Table~\ref{tab:param} summarizes the notations used in \method. More formally the modified equations are as follows: 
\begin{align}
    &\frac{dS_{i}}{dt} = - \frac{\beta_i S_{i}I_{i}}{N_{i}}\\
   &\frac{dE_{i}}{dt} = \frac{\beta_i S_{i}I_{i}}{N_{i}} - \sigma E_{i} + {\sum_{i} E_{ji} - \sum_{i} E_{ij}}\\
    &\frac{dI_{i}}{dt} = \sigma E_{i} - \gamma I_{i}\\
    &\frac{dR_{i}}{dt} = \gamma I_{i}
\end{align}

where $S_{i}$, $E_{i}$, $I_{i}$, $R_{i}$ represent the size of susceptible, exposed, infectious and recovered population in target population $i$. The incubation rate~\(\sigma\) is set to be \(\frac{1}{C}\) where C is the mean latent period of the disease while the recovery rate~\(\gamma\) is set to be \(\frac{1}{D}\) where D is the mean infectious period. The transmission rate~\(\beta_i\) encapsulates both the transmission rate of the disease and the contact rate of population $i$. We define the population reproduction number \popri as the basic reproduction number \popr of population $i$ and compute it as $\frac{\beta_i}{\gamma}$ to measure the average number of infections caused by infectious individuals of the same population, $I_{i}$.

$E_{ij}$ are the number of exposed individuals travelling from country $i$ to country $j$ and $E_{ji}$ is the number of individuals travelling from country $j$ to country $i$. As the net flow of exposed individuals is insignificant compared to the total population $N_i$, we kept $N_i$ unchanged throughout our simulations. $E_{ij}$ and $E_{ji}$ are estimated from the flight network as follows: 
\begin{align} 
E_{ij} &=  Passengers_{ij} \times P_{i} \times \eta\\
E_{ji} &=  Passengers_{ji} \times P_{j} \times \eta\ 
\end{align}
where $P_i$ denotes the test positive rate of country $i$, measured in the number of confirmed cases per thousand tests. $\eta$ denotes the percentage of exposed individuals over all infected individuals. $Passengers_{ij}$ and $Passengers_{ji}$ estimate the total number of passenger between population $i$ and $j$ and are estimated as:
\begin{align} 
Passengers_{ij} &=  F_{ij} \times CAP \times LF\\
Passengers_{ji} &=  F_{ji} \times CAP \times LF\
\end{align}

Here, \(F_{ij}\) and \(F_{ji}\) are the number of outbound/inbound flights from/to country $i$ to/from country $j$. $CAP$ gives an estimate for the average flight capacity. $LF$ proxies the passenger load factor (onboard passengers to available seats ratio) and offers the flexibility of modelling the reduction in passengers per flight due to the pandemic~\cite{aaimpact, itaaircap}. 

Why do we estimate the number of exposed individuals going in/out of country $i$, $E_{ji}$/$E_{ij}$ based on the test positive rate $P$ instead of the testing rate? Each country has different testing capacities and strategies. The number of confirmed cases alone does not necessarily reflect the actual number of infected individuals in each country~\cite{krantz2020level}. Test positive rate $P$, on the other hand, gives us a rough idea how many infected individuals we would find if we were to test everyone on the plane. The product of $Passengers$ and test positive rate $P$ only tells us how many infectious individuals would have come in or out of the population, assuming they have not been prevented from travelling. To derive the number of exposed individuals, we further multiply this with $\eta$, the percentage of exposed individuals over all infected individuals which include the exposed and the infectious. The same $\eta$ is used across populations.

\section{Estimating mobility between populations from real time data}
While \method can be used for modelling the spread of disease over multiple populations, in this paper \textit{we focus on one target population, \textit{Canada}}, to provide more straightforward comparison with the standard SEIR model, as well as to make collecting and estimating relevant data more manageable.


\begin{figure}[t]
    \includegraphics[width=0.99\textwidth]{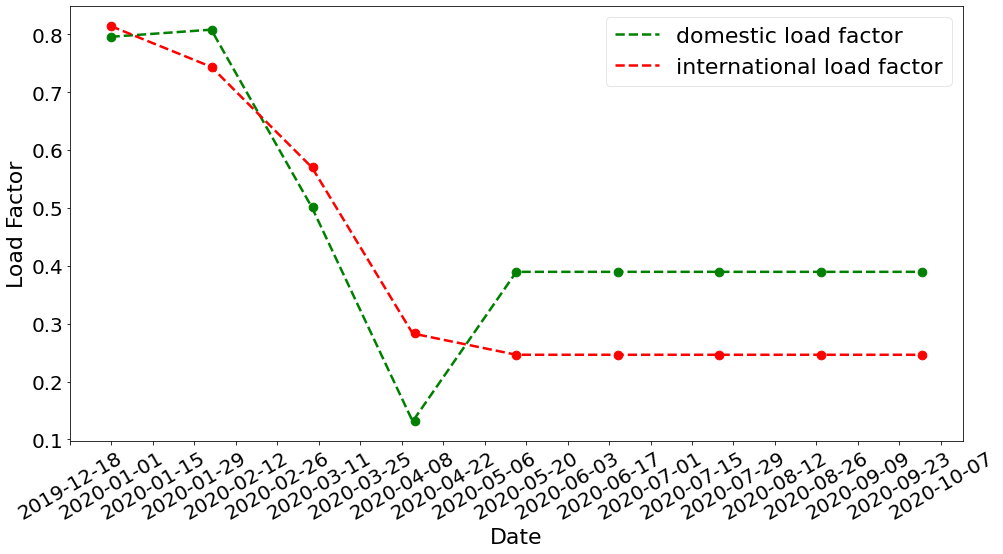}
    \caption{\csentence{Ratio of passengers to available seats (LF) in flights} \\interpolated based on the available data from \cite{ustransport}}. 
    \label{fig:load_factor}
\end{figure}

\begin{figure}[h!]
     \centering
        \begin{subfigure}[b]{.98\textwidth}
        \includegraphics[width=1\textwidth]{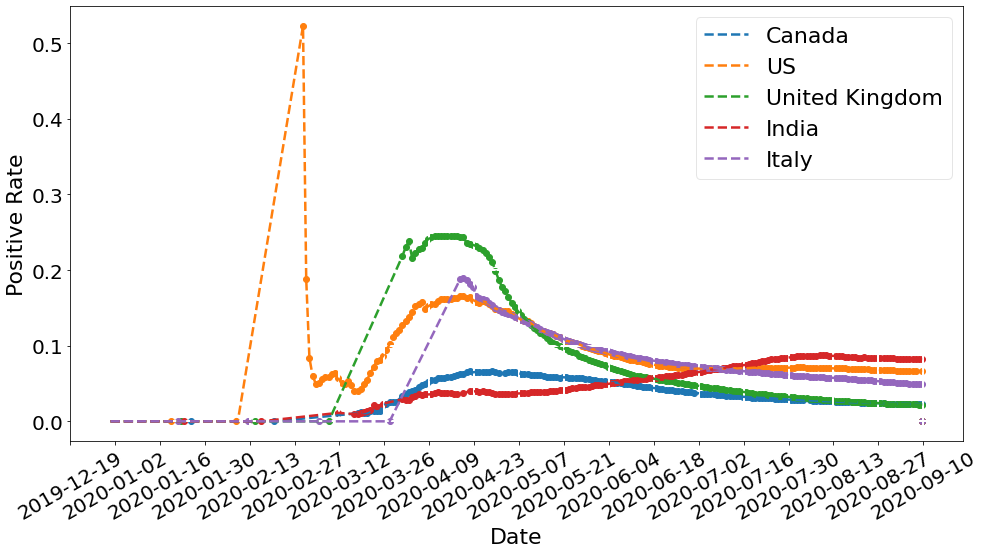}
        \caption{Test positive rates per country}   
      \end{subfigure}
    \begin{subfigure}[b]{.98\textwidth}
        \includegraphics[width=1\textwidth]{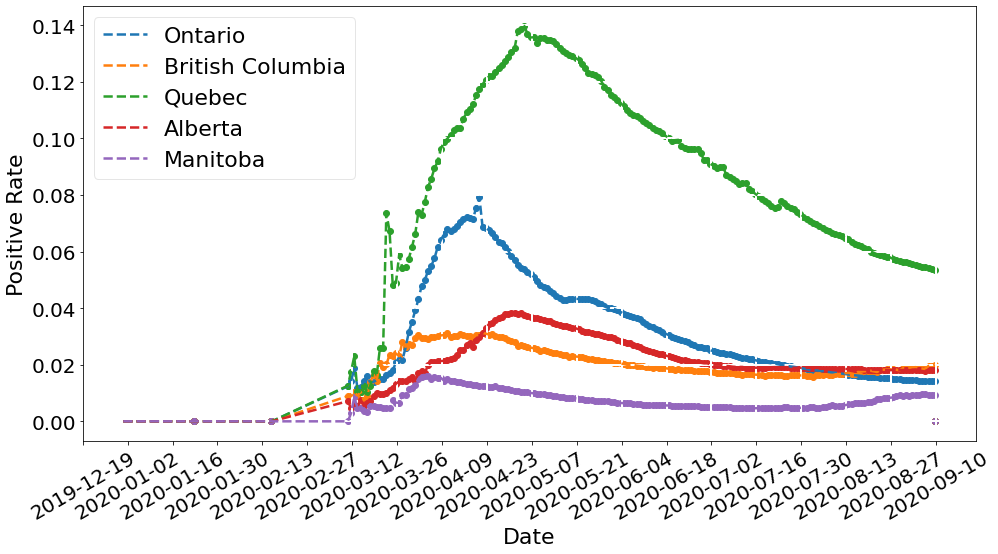}
        \caption{Test positive rates per Canadian province}
      \end{subfigure}
    \caption{\csentence{Test positive rates}, interpolated based on data from \cite{posrate, CanCovid}}
    \label{fig:pos_rate}
\end{figure}

We began our study few months after the pandemic started. Unlike accessing the real-time flight network, \textit{collecting historic flights and estimating population movements is a challenging task}. In this work, we made a few assumptions and we tried our best to validate our estimations against external sources. We believe that \method can be highly informative for decision making at public agencies which have access to more detailed and accurate flights data. We haven't been able to secure access to such data. Therefore, our experiments represent our best attempt at modelling \covid and serve as a proof of concept for the potentials of incorporating the flight network. 

To estimate $F_{ij}$ and $F_{ji}$, we collected historic flights data dating back to Jan, 2020~\cite{frd24}. To correct for possible missing flights, we scaled the data so that our account of monthly flights matches the domestic and international itinerant movements reported by Statistics Canada~\cite{statcaniter}. Based on the flights data, we constructed one snapshot of flight network per day. The nodes are the airports and the edges represent flight connections. They are directed and weighted by the number of flights on that day. 
As an example,  Figure~\ref{fig:fn1} and Figure~\ref{fig:fn2} shows the comparison between the flight network on January 20, 2020 and April 2, 2020. The size of the bubble is proportional to the air traffic at the airport. The width of the edge is in proportion to the number of flights between two airports. We observe that there are far fewer flights on April 2 compared to January 20. This can be partially attributed to the travel ban put in place in March~\cite{acapsgmd}. It can also be sentiment-related i.e. people are afraid to travel due to the pandemic. To some extent, this dynamic flight network reflects the impact of travel restrictions. 

\begin{figure}[h!]
    \includegraphics[width=0.99\textwidth]{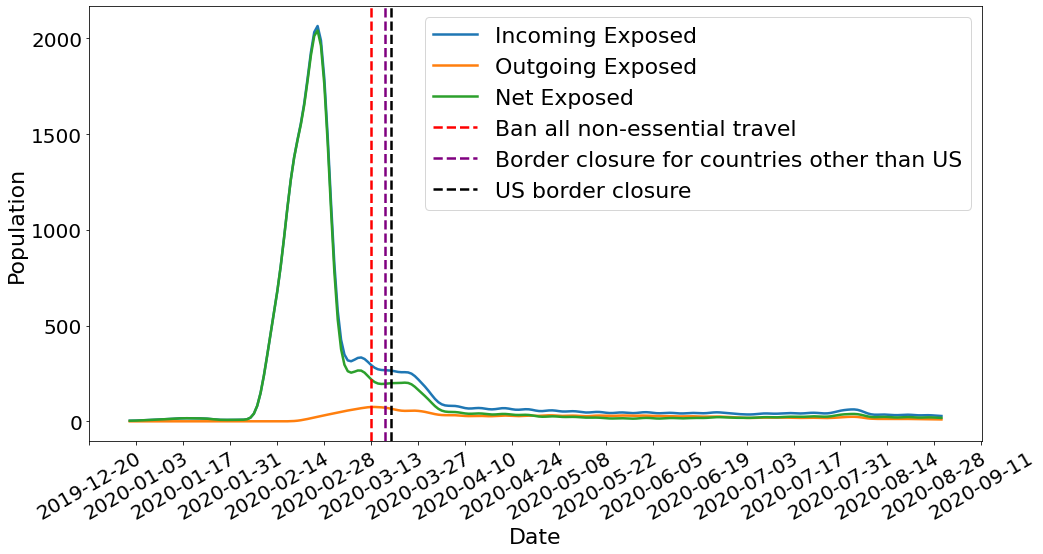}
    \caption{\csentence{The estimated number of exposed individuals going in and \\ out of Canada, before and after travel restrictions}}
    \label{fig:exposed_can}
\end{figure}

We then further map the airports to countries or provinces and aggregate the number of flights which serve as the input to \method. US passengers, flights and load factor data~\cite{ustransport} is used as a proxy to the flight capacity and the load factor since this information is not available for Canada. We compute the average flight capacity as:
\begin{equation}
    CAP = \frac{Passengers}{Flights \times Load Factor}
\end{equation}
$CAP$ is estimated to be 111 for international flights and 180 for domestic flights. Figure~\ref{fig:load_factor} shows the interpolated load factor $LF$. We can see that both domestic and international load factor have dropped significantly since the start of the year. At the time of writing, This data is only available from January to May. We assume that the load factor remains the same from May onward. 

\begin{figure}[h!]
     \centering
        \begin{subfigure}[b]{.98\textwidth}
        \includegraphics[width=1\textwidth]{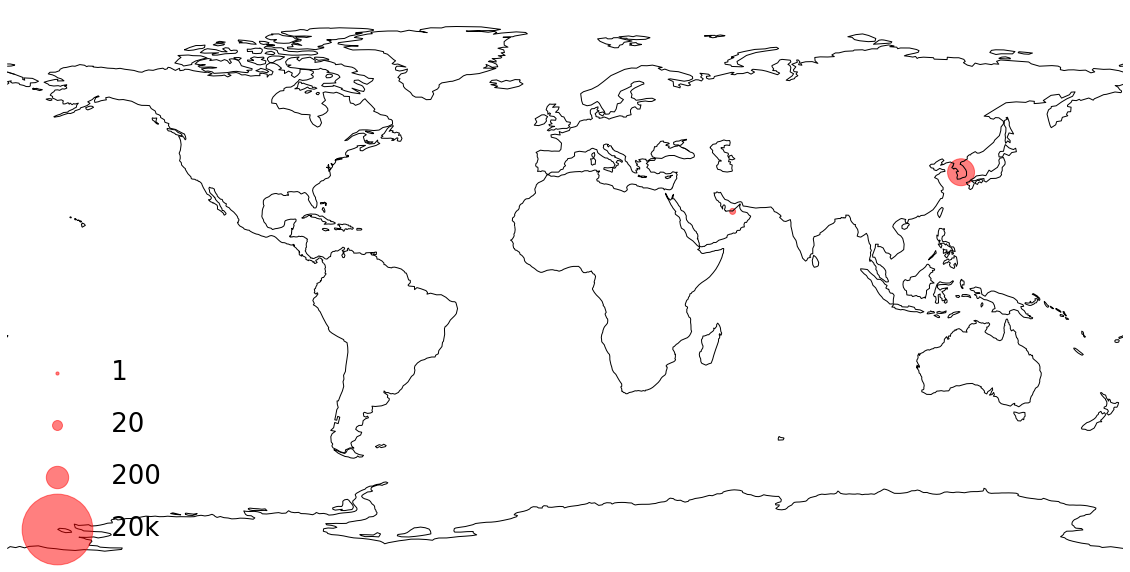}
        \caption{\csentence{Total incoming exposed individuals by January 20th, 2020} }   \label{fig:exp1}
      \end{subfigure}
    \begin{subfigure}[b]{.98\textwidth}
        \includegraphics[width=1\textwidth]{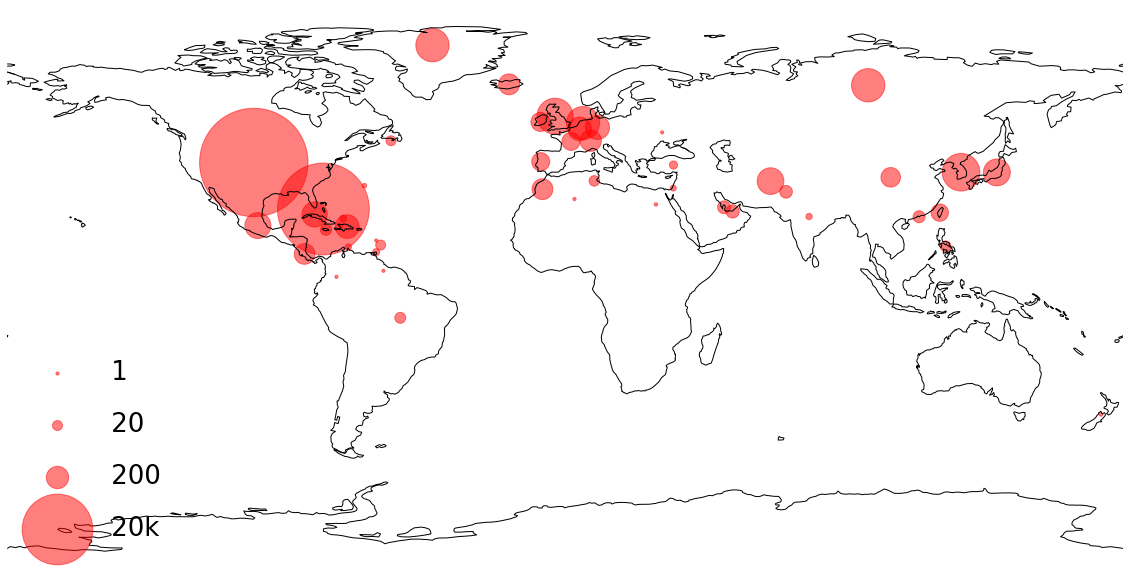}
        \caption{\csentence{Total incoming exposed individuals by April 2nd, 2020}}\label{fig:exp2}
      \end{subfigure}
 \caption{\csentence{Total incoming exposed individuals to Canada aggregated \\ per country}. Note that the circle sizes are plotted in log scale.}
     \label{fig:exposedOverview}
\end{figure}

We obtain country-level test positive rates from public dataset~\cite{posrate} published by Oxford Martin School. For provinces and territories within Canada, we use data published by the Government of Canada~\cite{CanCovid}. Figure~\ref{fig:pos_rate} shows the interpolated test positive rates for selected countries and provinces. $\eta$ is set to 10\% to roughly agree with the tracked travel-related (imported) cases reported by \cite{berry2020open}. 
Figure~\ref{fig:exposed_can} shows the estimated movements of exposed people to and from Canada. We observe a spike of incoming exposed individuals in February. The estimation drops after the various travel-related NPIs were put in place. The number of outgoing exposed individuals are low because Canada has a comparatively low test positive rate when compared to the rest of the world. Figure~\ref{fig:exposedOverview} shows our estimation of total incoming exposed individuals to Canada by January 20 and April 2.

\paragraph{Estimating unrestricted flight volume:} We assume the January's data to be representative of the air traffic prior to COVID-19. Therefore, we scaled the air traffic in the following months to match January's volume to simulate demographic dynamics without travel restrictions in the early pandemic detection simulations. Moreover, we used the same scaling procedure to simulate the air traffic when we lift travel restrictions and used a scaling factor $\tau$ to control the amount of flights we resume, in the corresponding simulations discussed below. Given that travel restrictions affect both the number of flights and the number of passengers the plane, we also scale the load factor to account for changes in passengers per flight caused by the relaxation of travel restrictions.  The same scaling factor $\tau$ is used to scale both air traffic volume and load factor. 

\section{Simulations and Discussions}

In the following sections, we describe how \method can be applied to \emph{1)} predict uncontrolled early time epidemic dynamics, \emph{2)} assess the impact of travel restrictions, \emph{3)} estimate the basic reproduction number \popr and \emph{4)} simulate the effect of lifting travel restrictions. 
For all our experiments, we adopted mean latent period $C$ to be 5 and mean infectious period $D$ to be 14~\cite{whocjm} suggested by WHO. The transmission rate \(\beta\) is obtained by fitting \method to country-specific COVID-19 dataset published by the government of Canada~\cite{CanCovid, CanPop}. 

\subsection{Predict uncontrolled early time epidemic dynamics}
\label{sub:noRestrict}

\begin{figure}[t]
    \includegraphics[width=0.98\textwidth]{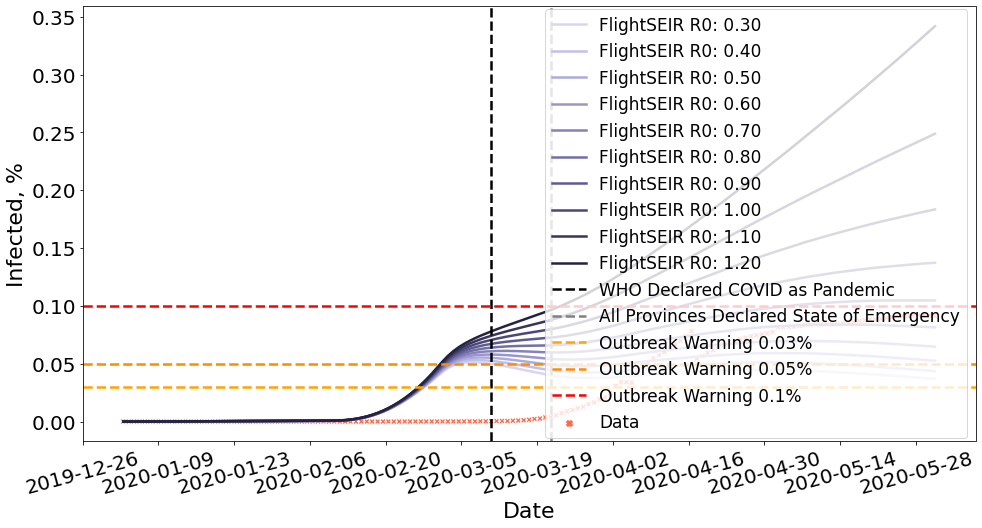}
    \caption{\csentence{Early time prediction \emph{without} enforced travel restriction}}
    \label{fig:etpCanadawotr}
\end{figure}
In the early stage of an outbreak, the majority of confirmed cases are imported from outside populations. By monitoring flights from countries with confirmed cases, \method can estimate the risk of an outbreak caused by the inflow of pre-symptomatic cases and provide timely signals to the need for travel restrictions and contact tracing. In particular, \method can provide an alert when the size of an outbreak exceeds a given threshold. In our simulations, we consider an outbreak occurs when the percentage of infected cases in the total population exceeds 0.05\%. This is in accordance with the World Health Organization's definition of a disease outbreak~(the occurrence of disease cases in excess of normal expectancy)~\cite{outbreakdef}.

To predict early time dynamics without travel restrictions, we run \method on flights network generated with a scaling factor of $\tau=1$, simulating the air traffic prior to the pandemic. \popr used for this experiment can be chosen based on existing research on the pathogen and can be adjusted as more information becomes available. In this paper, we consider a range of [0.3, 1.2] for \popr. Many works~\cite{verity2020estimates, abdollahi2020temporal} reasoned that early estimation of $R0$ is not accurate due to reasons such as lack of testing capabilities. Therefore, we experiment with a wider range of parameters to include both the best and worst case scenarios. The simulation runs from January 2nd, when there is no confirmed cases, to June 1st, when the travel restrictions have taken full effects.

\begin{observation}
    When imported cases from outside the population are incorporated into the model, even \popr$< 1$ can lead to an outbreak.
\end{observation}

Figure \ref{fig:etpCanadawtr} shows the early time prediction for Canada, assuming no travel restriction is imposed. As shown in the figure, more than 0.053\% of the population will be infected by June 1st if \popr$\geq 0.5$. More than 0.104\% of the population will contract the disease if \popr$\geq 0.8$. Moreover, the number of cases is still growing by the time the simulation ends. This is in contrast with the belief in traditional epidemiology that the disease will die out when \popr$< 1$~\cite{barabasi2013network}. This seemingly contradicting result can be explained by the different assumptions made by these two models. The standard SEIR model assumes that the population is a closed system where individuals do not move between the populations. However, this is often unrealistic considering the huge amount of domestic and international flights annually. On the other hand, \method considers demographic dynamics between populations. As a result, even if each infectious individual infects less than one susceptible person, the disease can still persist because exposed individuals from other populations can move into the population of interest and transition to the infectious compartment.

\subsection{Assess the impact of travel restrictions}

To estimate the real impact of travel restrictions, we run \method with real flights data and the same set of \popr. By comparing the results for this experiment with those in Section~\ref{sub:noRestrict}, we can investigate the effect of reducing air traffic on the spread of the disease. Table \ref{table:earlypred} further quantifies the effect in terms of the different in the infected individuals in both senarios. 

\begin{observation}
    Travel-related NPIs have a significant impact on the spread of the disease. 
\end{observation}

\begin{figure}[t]
    \includegraphics[width=0.98\textwidth]{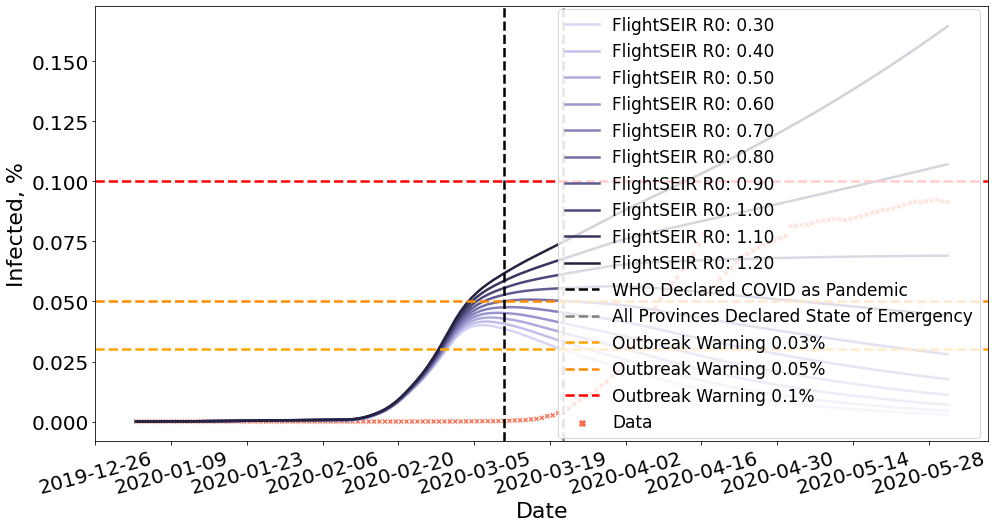}
    \caption{\csentence{Early time prediction \emph{with} enforced travel restriction}} 
    \label{fig:etpCanadawtr}
\end{figure}

\begin{table}[h!]
      \begin{tabular}{cccc}
        \hline
        \popr & Without Travel Res.  & With Travel Res.  & Difference\\ \hline
        0.3   & 0.037\%              & 0.003\%           & 0.034\% \\
        0.4   & 0.044\%              & 0.004\%           & 0.039\% \\
        0.5   & 0.053\%              & 0.007\%           & 0.046\% \\
        0.6   & 0.065\%              & 0.011\%           & 0.054\% \\
        0.7   & 0.081\%              & 0.018\%           & 0.064\% \\
        0.8   & 0.105\%              & 0.028\%           & 0.077\% \\
        0.9   & 0.137\%              & 0.044\%           & 0.093\% \\
        1.0   & 0.183\%              & 0.069\%           & 0.114\% \\
        1.1   & 0.249\%              & 0.108\%           & 0.142\% \\ 
        1.2   & 0.342\%              & 0.165\%           & 0.177\% \\ \hline
      \end{tabular}
      \caption{Estimated percentage of infected people with and without travel restrictions on June 1st.}
      \label{table:earlypred}
\end{table}

Figure~\ref{fig:etpCanadawtr} projects the spread of disease in Canada with travel restriction and table~\ref{table:earlypred} shows the results in percentage. We can see that imposing travel restrictions can reduce the percentage of infectious individuals by up to 0.177\% of the total population. Moreover, the impact of travel restriction increases with \popr. With \popr$=0.5$, the infected population remains stable at 0.053\% without travel restriction whereas the curve peaks in late February and drops to 0.007\% with travel restriction. With \popr$=1.2$, the infected population is estimated to reach 0.342\% if the government had not issued the travel restrictions. In the meantime, only 0.165\% of population will be infected if the measure is in place. 

\subsection{Estimate the basic reproduction number}

\method offers a way to decouple external source of infection with community transmission and provides a more accurate estimation of \popri (the average number of infections caused by infectious individuals of the same population, $I_{i}$). For brevity, we will drop the subscript $i$ and use \popr instead. \method can also infer the initial size of infected population from the dynamic flight network whereas the standard SEIR model requires at least one exposed or infected individual to be present in the population at $t=0$.

To estimate \popr, we run \method with real air traffic and search through a set of \popr. We select the one with the least mean squared error (MSE) compared to the confirmed cases. When estimating \popr, one challenge is that the data we are fitting to does not necessarily reflect the reality due to insufficient testing~\cite{verity2020estimates, abdollahi2020temporal}. \method almost always predicts more than what was reported but it does not mean that \method overestimates the number of infected people. To account for this, we compute MSE only for time steps where test rate exceed certain threshold. To provide a point for comparison, we also run the standard SEIR model and fit \popr with the same loss function. 

\begin{figure}[t]
    \includegraphics[width=0.98\textwidth]{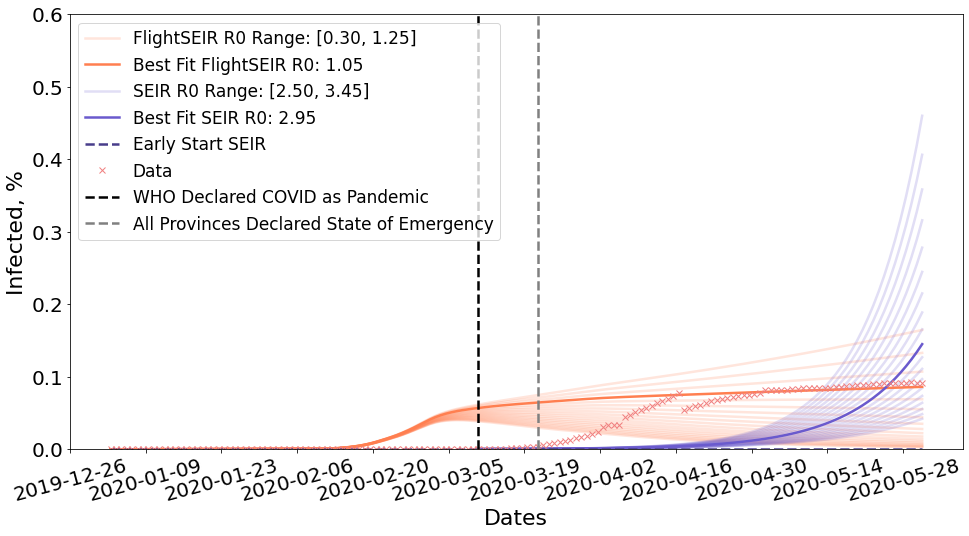}
    \caption{\csentence{\popr Estimation for Canada.} Initially, \method predicts more \\than the confirmed cases while the standard SEIR model underestimates the \\ number of infectious individuals. Later on, \method fits almost perfectly to \\ the confirmed cases whereas the standard SEIR model starts to overestimate.}    \label{fig:betafit1}
\end{figure}
Figure~\ref{fig:betafit1} shows the estimation of \popr given by \method and the standard SEIR model for Canada. To illustrate what happens if we start SEIR before there are any confirmed cases, we run Early Start SEIR (the dashed purple line) for which we set $E0$ and $I0$ both to 0. We see that its prediction remains a flat line throughout the experiment. We start another SEIR model (the solid purple line) the date on which total confirmed cases in Canada exceed 100 people. We search through \popr range of [2.50, 3.45] and find the best fit to be 2.95. We observe that, at the beginning, the standard SEIR model underestimates the number of infectious individuals compared to the confirmed cases. However, its prediction grows exponentially and shows signs of overestimation by the end of May. For \method, the \popr range is set to be between 0.30 and 1.25 and the best fit is estimated to be 1.05. \method is predicting more infectious individuals than the confirmed cases up until late April but its curve fits almost perfectly afterwards.

\begin{observation}
    \popr predicted by the standard SEIR model is significantly higher than the one estimated by \method.
\end{observation}

The best fit \popr for the standard SEIR model is 2.95 whereas the best fit for \method is 1.05. One explanation is that the standard SEIR model assumes that the population is closed and all infections come from within. As a result, it requires a high \popr to sustain a fast initial growth. On the other hand, \method acknowledges that there can be exposed individuals coming from outside the population and therefore requires a lower rate for community transmission. 

\begin{observation}
    \method predicts more infected individuals than the data at the beginning of the outbreak and the difference between the two approximates the degree of under reporting.
\end{observation}

\method predicts that there were 21620 infected individuals in Canada by the time WHO declared COVID-19 as pandemic on March 11. In the meantime, Canadian government only reported 100 confirmed cases~\cite{CanCovid}, an under-reporting of 21520\%. By the time all provinces or territories in Canada declared a state of emergency or a public health emergency on March 22, \method estimates 24465 people to have been infected while only 1438 cases were confirmed, an under-reporting of 1601\%. The discrepancy between our model prediction and official report can be attributed to insufficient testing. From March 22 to June 1st, the test rate in Canada went up by 1612\% from 2.618  (tests per thousand) to 44.812(tests per thousand)~\cite{posrate} while the percentage of infected population went up by 2309\% from 1438 to 34640 people. On June 1st, \method's prediction and the data only differ by 6\%.

\begin{figure}[h!]
     \centering
        \begin{subfigure}[b]{.98\textwidth}
        \includegraphics[width=\textwidth]{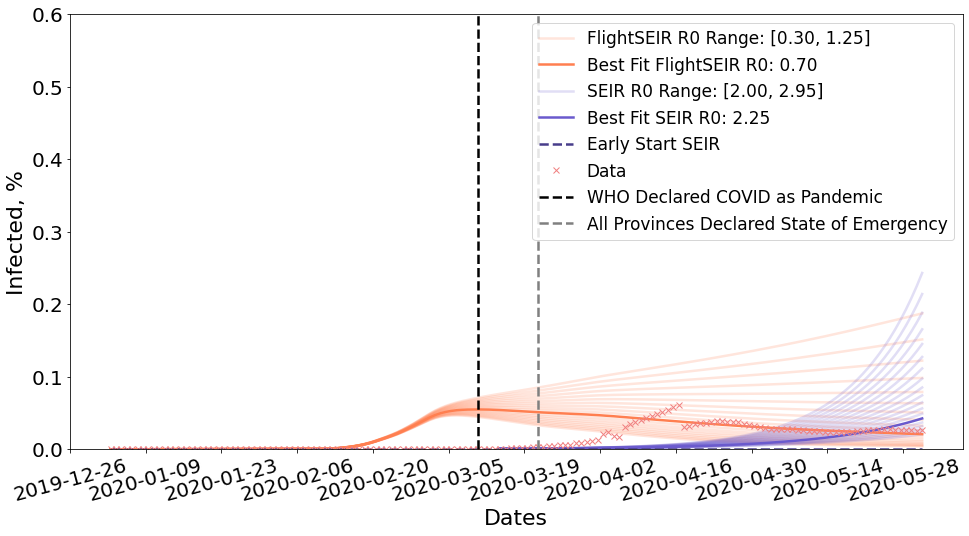}
        \caption{\csentence{Ontario}}   
      \end{subfigure}
    \begin{subfigure}[b]{.98\textwidth}
        \includegraphics[width=\textwidth]{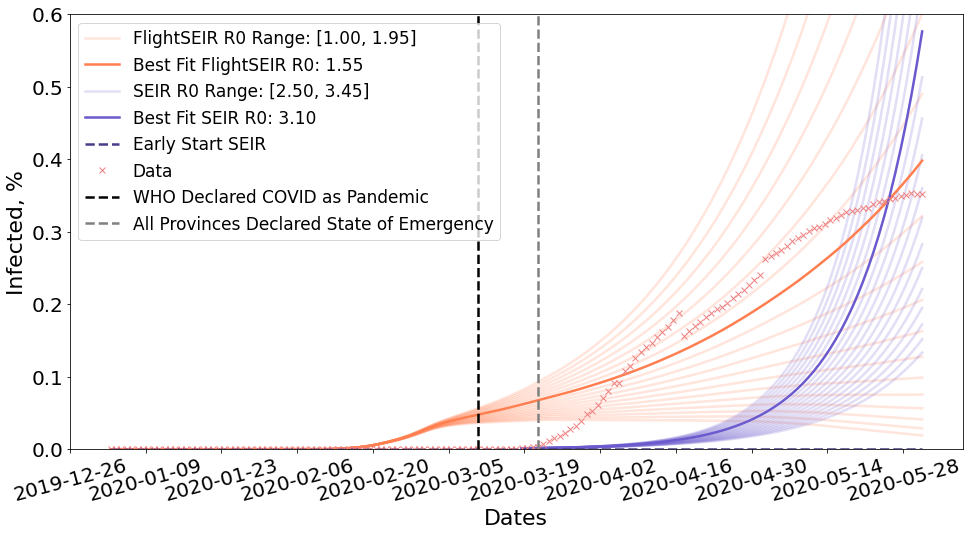}
        \caption{\csentence{Quebec}}
      \end{subfigure}
    \caption{\csentence{\popr Estimation for difference provinces within Canada.} \\ Quebec is estimated to have a much higher \popr than Ontario. In both cases, \\ \method shows a better fit than the standard SEIR model.}
    \label{fig:betafit2}
\end{figure}

Figure~\ref{fig:betafit2} shows the fitting for different provinces and territories within Canada. In almost all cases, \method demonstrates a better fit than the standard SEIR model. Note that we limit the date range to be between January 2nd and June 1st because \method is only designed to account for travel-specific measures and does not consider later changes in \popr due other NPIs such as social distancing. 


\subsection{Evaluate the risks of lifting travel restrictions}

\begin{figure}[t]
    \centering
    \includegraphics[width=0.98\textwidth]{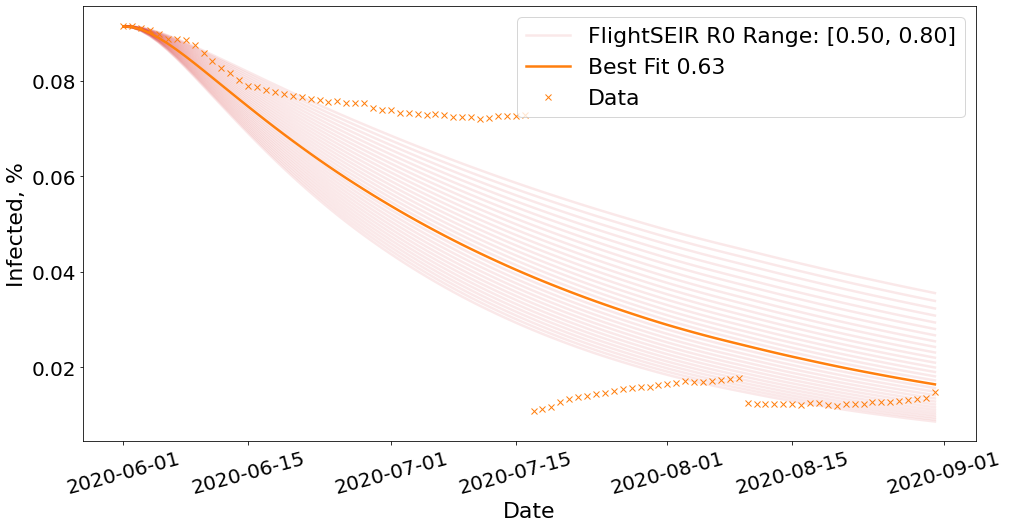}
    \caption{\csentence{\popr estimation for reopening simulation.} \method is fitted  \\ to confirmed cases from June 1st to August 1st, approximately 2 months before \\ the reopen date. We run grid search in the range of [0.50, 0.80] and find the \\ best fit to be 0.63.}
    \label{fig:reopen0_can}
\end{figure}

Next, we study the impact of lifting travel-related NPIs such as resuming flights and opening borders. We assume the reopen date to be August 1st and run \method with air traffic generated from various the scaling factor $\tau$ to simulate reopening at different scales. More specifically, we model the scenarios in which we resume 25\%, 50\%, 75\% and 100\% of air traffic. We obtain \popr by fitting \method to confirmed cases from June 1st to August 1st, approximately 2 months before the reopen date. We assume that it remains the same for the duration of the simulation. Figure~\ref{fig:reopen0_can} shows the fitting for Canada. We did grid search in the range of [0.50, 0.80] and found the best fit to be 0.63. We can tell from the curve that the number of cases already started to drop and the disease appears to have been contained.   

\begin{figure}[h!]
    \centering
    \begin{subfigure}[b]{.98\textwidth}
        \includegraphics[width=\textwidth]{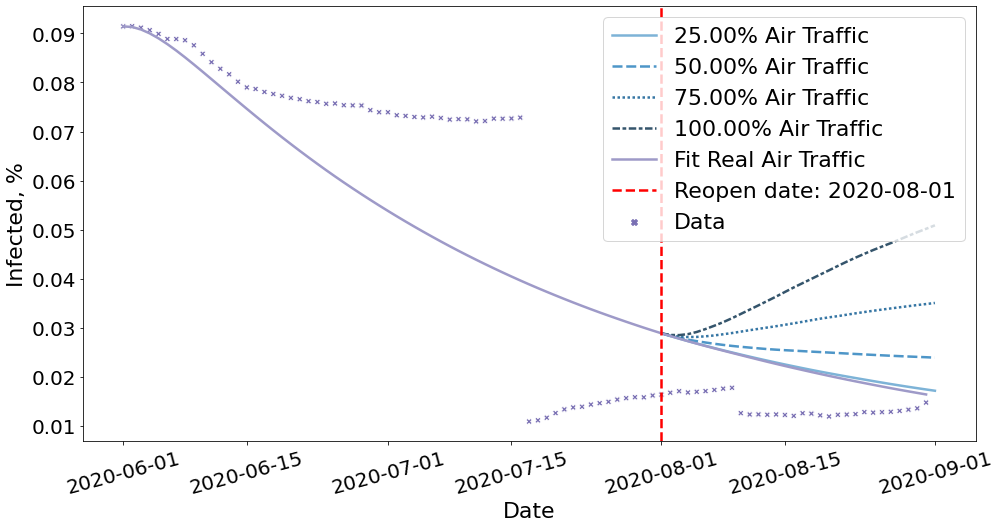}
        \caption{\csentence{Percentage of active infectious individuals}}   
    \end{subfigure}
    \begin{subfigure}[b]{.98\textwidth}
        \includegraphics[width=\textwidth]{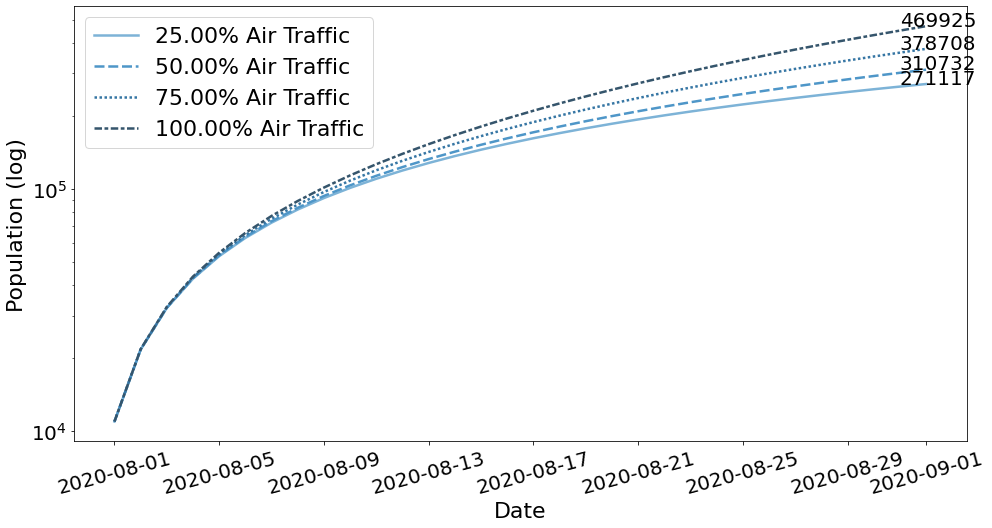}
        \caption{\csentence{Cumulative infections}}
    \end{subfigure}
    \caption{\csentence{Simulation of resuming international flights in Canada.}\\ The figure shows the effect of resuming 25\%, 50\%, 75\% and 100\% air traffic \\ between Canada and the rest of the world. The simulation starts on August 1st \\ and continues for a month. We observe an immediate rebound when flights \\ are increased by more than 50\%.}
    \label{fig:reopen1_world}
\end{figure}

Figure~\ref{fig:reopen1_world} and Table~\ref{table:reopen} show the projections for resuming flights to and from countries all over the world. In addition, we run \method with real air traffic and plot its prediction as well as the confirmed cases as a baseline. Running \method with real air traffic and 25\% air traffic produce similar curves, indicating that the real volume in August is only 1/4 of what it was before. When we resume flights for 50\% or more, we immediately see a rebound in the number of infectious people. The model shows that if we resume 100\% air traffic, we would have 469925 cumulative cases by September 1st compared to only 271117 infections if we remain at 25\%.

\begin{observation}
    Both the number of imported cases and community transmission increase with the scale of reopening.
\end{observation}

\begin{figure}[h!]
    \centering
    \includegraphics[width=0.98\textwidth]{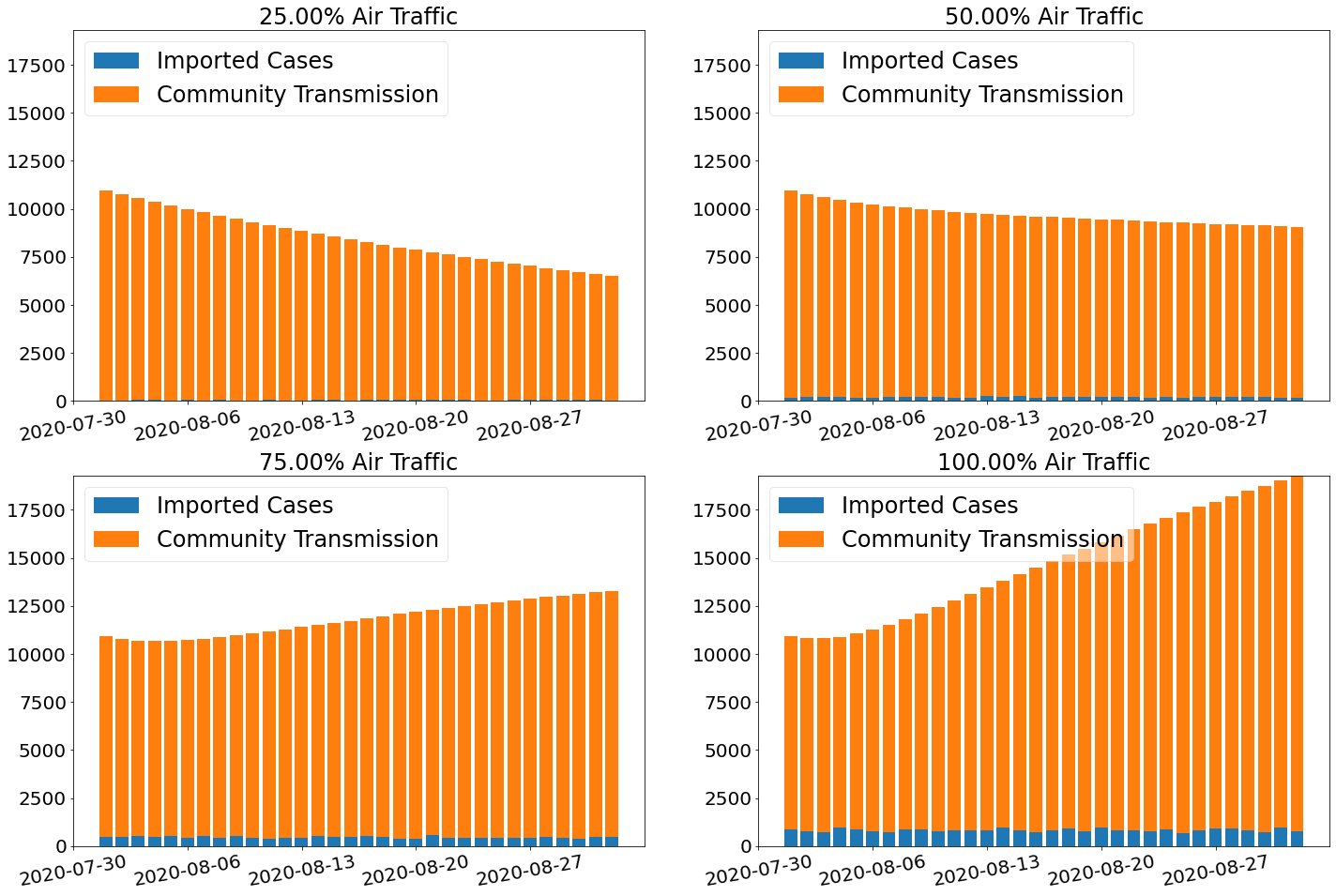}
    \caption{\csentence{Daily active cases by source of infection.} The figure shows \\ the composition of infected population if we resume 25\%, 50\%, 75\% and 100\% \\ of air traffic. Both the number of imported cases and community transmission \\ increase with the scale of reopening.}
    \label{fig:reopen4_world}
\end{figure}

Figure~\ref{fig:reopen4_world} and Table~\ref{table:reopenbysource} show the daily active cases break down by source of infection. The number of imported cases increase with the scale of reopening. \method predicts that by September 1st, there will be 43 imported cases per day if we only resume 25\% air traffic compared to 780 per day if we resume 100\% air traffic. \method expects there to be 6475 community transmission per day if we only resume 25\% air traffic as opposed to 18507  per day if we resume 100\% air traffic. For the same scaling factor $\tau$, the ratio of imported cases to community transmissions will decrease if we lift travel restrictions for a long period of time. From August 1st to September 1st, if we resume 100\% air traffic, the daily imported case will decrease from 865 to 780 whereas the daily infection caused by community transmission will increase from 10086 to 18507. The ratio of imported cases to community transmission decreases from 0.086 to 0.042. This may be because with the imported cases, we have a larger infected population base to infect other people and thus more community transmissions. 

\begin{table}[h!]
      \begin{tabular}{ccccc}
        \hline
        Scaling Factor $\tau$ & Imported & Community & Total & Imported/Community \\ \hline
        0.25               & 43       & 6475      & 6518  & 0.0066 \\
        0.50               & 176      & 8895      & 9071  & 0.0198 \\
        0.75               & 490      & 12803     & 13293 & 0.0383 \\
        1.00               & 780      & 18507     & 19287 & 0.0421 \\ \hline
      \end{tabular}
      \caption{Daily active cases by source of infection by September 1st if we resume flights between Canada and the rest of the world by 25\%, 50\%, 75\% and 100\%.}
      \label{table:reopenbysource}
\end{table}

\begin{observation}
    The effect of reopening depends on countries and we need to evaluate the risk of lifting travel restrictions on a case-by-case basis.
\end{observation}

\begin{figure}[h!]
    \centering
    \begin{subfigure}[b]{.98\textwidth}
        \includegraphics[width=\textwidth]{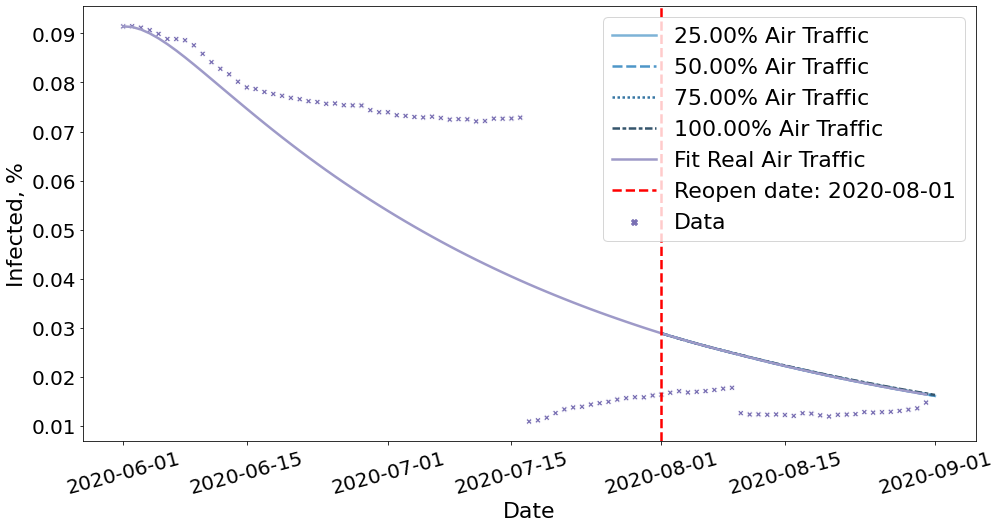}
        \caption{\csentence{Reopen between Canada and UK}}   
    \end{subfigure}
    \begin{subfigure}[b]{.98\textwidth}
        \includegraphics[width=\textwidth]{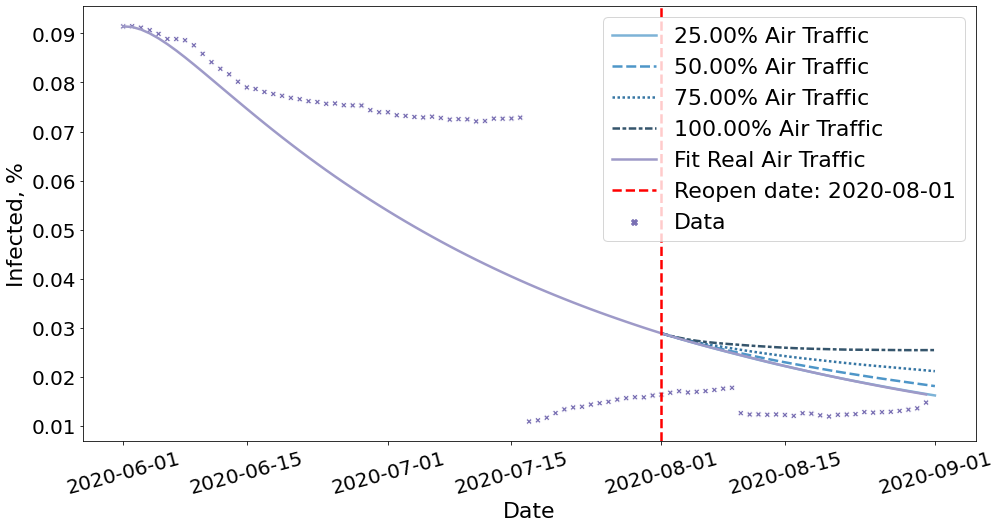}
        \caption{\csentence{Reopen between Canada and US}}
    \end{subfigure}
    \caption{\csentence{Reopening simulation with two different countries}.\\This shows the effect of resuming 25\%, 50\%, 75\% and 100\% air traffic between  \\Canada and UK/US. The impact of resuming flights with UK is negligible when \\ compared to that of US.}
    \label{fig:reopen1_others}
\end{figure}

Figure~\ref{fig:reopen1_others} and Table~\ref{table:reopen} show the results for reopening with the United States versus United Kingdom. We estimate that the cumulative cases will differ by 52387 people if we resume 100\% with the US as opposed to remain at current traffic. However, the cumulative cases will differ by 1239 for UK. The impact of resuming flights with UK is negligible compared to that of US. This can be explained by the fact that half of Canada's international flights are coming from the US and that UK has far lower positive rate than the US. 

\begin{table}[h!]
      \begin{tabular}{cccc}
        \hline
        Scaling Factor $\tau$ & All Countries  & US only  & UK only\\ \hline
        0.25               & 271117         & 266058   & 266011 \\
        0.50               & 310733         & 277027   & 266179 \\
        0.75               & 378708         & 294487   & 266722 \\
        1.00               & 469925         & 318445   & 267250 \\ \hline
      \end{tabular}
      \caption{Estimated Cumulative Infections by September 1st if we resume flights between Canada and other countries by 25\%, 50\%, 75\% and 100\%.}
      \label{table:reopen}
\end{table}

\begin{observation}
    Resuming flights have different implications for different provinces and territories.
\end{observation}

\begin{figure}[h!]
    \centering
    \begin{subfigure}[b]{.98\textwidth}
        \includegraphics[width=\textwidth]{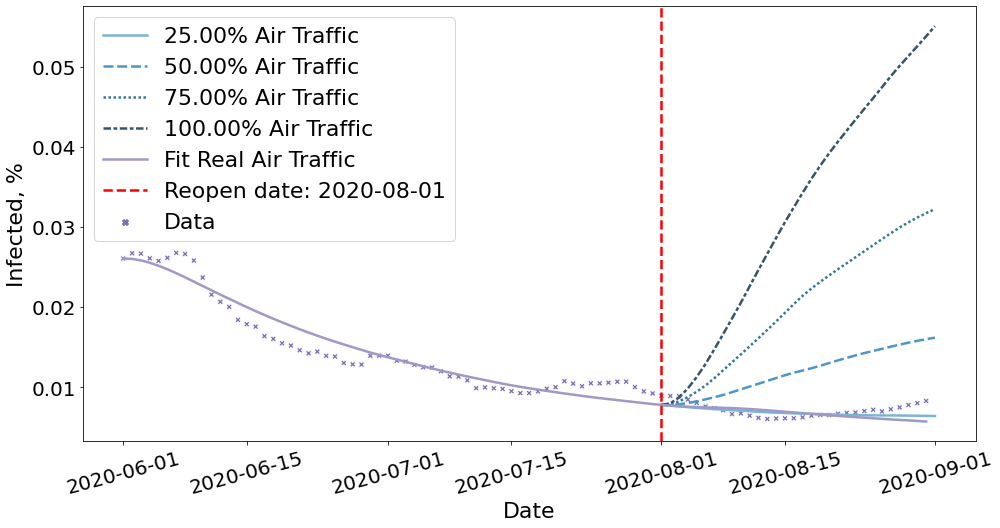}
        \caption{\csentence{Reopen between Ontario and the rest of the world}} 
    \end{subfigure}
    \begin{subfigure}[b]{.98\textwidth}
        \includegraphics[width=\textwidth]{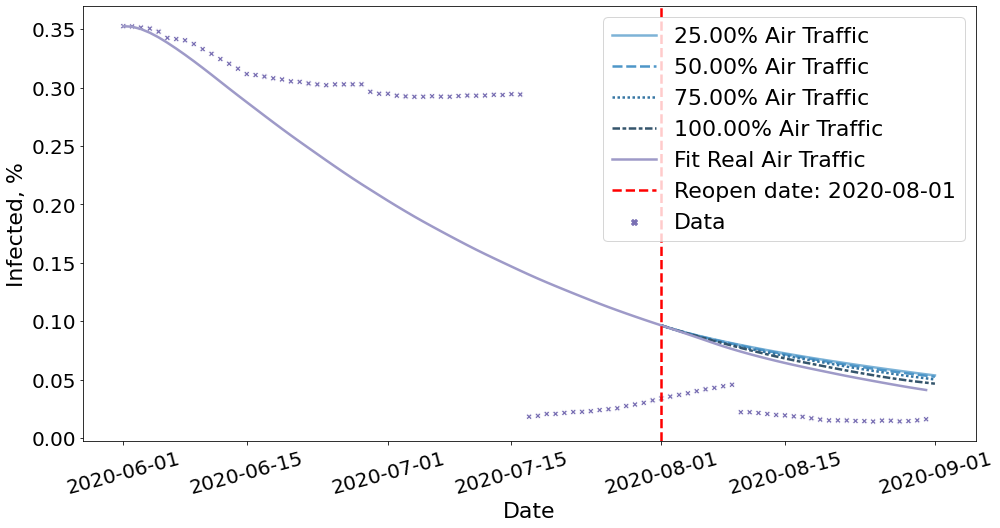}
        \caption{\csentence{Reopen between Quebec and the rest of the world}}
    \end{subfigure}
    \caption{\csentence{Reopening simulation for two provinces within Canada.} \\ The figure shows the effect of resuming 25\%, 50\%, 75\% and 100\% air traffic \\ between Ontario/Quebec and the rest of the world. While lifting travel \\ restriction is expected to have a mild impact on Quebec, we observe an \\ immediate rebound upon reopening Ontario.}
    \label{fig:reopen1_prov}
\end{figure}

Figure \ref{fig:reopen1_prov} shows the estimated risks of reopening for Ontario and Quebec. While both experiments simulate the scenario in which we resume flights to and from all countries, the impact is different for Ontario and Quebec. Ontario is estimated to have far worse rebound than Quebec. This may be due to the fact that Ontario have more international flights than Quebec in the dataset we collected. Another reason may be that \method consider the network flow of exposed individuals i.e. the difference between incoming and outgoing exposed. As Quebec has higher positive rate than Ontario as shown in Figure~\ref{fig:pos_rate}, it is estimated to have much more exposed individuals leaving the province when we resume flights. Therefore, even if both provinces receive the same amount of incoming exposed people, Ontario would be at greater risks.

\section{Conclusion}
In this work, we propose a modification to the widely used SEIR model, to derive inflow and outflow of exposed individuals from flight information. Our proposed \method is better suited for modelling the spread of the disease in a global pandemic such as COVID-19. The main contributions of \method are three-folds:
\begin{itemize}
    \item Enables early detection of outbreaks by taking into consideration the demographic dynamics of the population.
    \item Provides a more accurate estimation of the parameters, in particular the reproduction number \popr, and therefore facilitates a better understanding of the disease.
    \item Simulates the impact of travel restriction and evaluates the implications of lifting these measures.
\end{itemize}
Even though the flight network is well recorded, access to it is still restricted. We are working towards securing access to more accurate travel records to tune our estimations. We would also like to use \method for modelling the spread of disease in multiple populations simultaneously. In the multi-population setting, each country or node in the flight network will have its own SEIR model and the inter-population dynamics are proxied by the flight connections. We believe that this should be the modelling used when facing a global pandemic.  

\begin{backmatter}

\section{Declarations}

\subsection{Availability of data and materials}

The data for flight network is available from the corresponding author upon request.

The code will be released upon final submission and will be available at: https://github.com/CharlotteXiaoYeDing/FlightSEIR

All other datasets are publicly available:
\begin{itemize}
    \item Canada \covid case data, test positive rate $P$: https://www.canada.ca/en/public-health/services/diseases/2019-novel-coronavirus-infection.html\#a1
    \item Flights statistics: https://www150.statcan.gc.ca/t1/tbl1/en/tv.action?pid=2310000801 A
    \item Population $N_i$: https://www150.statcan.gc.ca/t1/tbl1/en/tv.action?pid=1710000901
    \item Load factor $LF$ and flights capacity $CAP$: https://www.transtats.bts.gov/Data\_Elements.aspx?Data=5 
    \item Global test positive rates: https://ourworldindata.org/coronavirus-testing 
\end{itemize}

\subsection{Competing interests}
The authors declare that they have no competing interests.
  
\subsection{Funding}
This research is funded by CIFAR Pan-Canadian AI Strategy.

\subsection{Authors' contributions}
XD and SH extracted flight network data. All authors designed the experiments. XD implemented the model and conducted the experiments. All authors participated in the writing of the manuscript.



\bibliographystyle{bmc-mathphys} 
\bibliography{bmc_article}      

\end{backmatter}
\end{document}